\documentclass[11pt]{article}
\usepackage{moriond,epsfig}

\def\Journal#1#2#3#4{{#1} {\bf #2}, #3 (#4)}

\def\PRL{\em Phys. Rev. Lett.}
\def\PRD{{\em Phys. Rev.} D}

\def\be{\begin{equation}}
\def\ee{\end{equation}}
\def\bea{\begin{eqnarray}}
\def\eea{\end{eqnarray}}

\begin{document}
\vspace*{4cm}
\title{Di-electron Continuum at PHENIX}

\author{ Y.L.~Yamaguchi for the PHENIX collaboration}

\address{Center for Nuclear Science, Graduate School of Science, University of Tokyo, \\ 
7-3-1 Hongo, Bunkyo, Tokyo, 113-0033, Japan}

\maketitle\abstracts{
The PHENIX experiment at RHIC has been carried out to investigate the properties of QGP.
Di-electron yields in p+p and Au+Au collisions have been measured.
An enhancement of the yield over a hadronic cocktail calculation is clearly seen for 
$p_{T} < 1.0~$GeV/$c$ and $150 < m_{ee} < 750~$MeV/$c^{2}$ in Au+Au collisions,
while the result in p+p collisions is consistent with the calculation.
The fraction of the virtual direct photon component to the di-electron yield is 
measured from a shape analysis using the di-electron mass distributions for 
$1.0 < p_{T} < 5.0~$GeV/$c$ and $m_{ee} < 300~$MeV/$c^{2}$, and the real direct photon 
spectra are deduced from the fractions obtained for p+p and Au+Au collisions. 
An excess of the direct photon yield above a binary-scaled p+p result is seen in Au+Au 
collisions.
The excess is fitted with an exponential function with an inverse slope parameter of 
$221 \pm 23 \pm 18~$MeV.
}

\section{Introduction}
The Quark Gluon Plasma (QGP) is a de-confined phase of a strongly-interacting matter.
The lattice QCD calculation predicts a phase transition from a hadronic phase to QGP 
at a critical temperature of $150 \sim 200~$MeV.
The study of QGP is very important since it gives not only new insights and intimate 
understanding of the strong interaction but also the understanding of the early universe 
at several $\mu$s after the Big Bang.

Heavy ion collision provides a unique tool to investigate the properties of QGP.
A lot of very exciting results such as large energy loss of light quarks, gluons
~\cite{PRL1} and heavy quarks~\cite{PRL2} and the strong elliptic flow for various 
particles~\cite{PRL2}$^{,}$~\cite{PRL3} has been reported for $\sqrt{s_{NN}} = 200~$GeV Au+Au 
collisions at Relativistic Heavy Ion Collider (RHIC).
These experimental facts indicate that the created matter has high-density and 
is rapidly-thermalized.

Di-leptons and direct photons are important probes to study QGP.
They are emitted through all the stages of the collisions, and penetrate the 
strongly-interacting matter.
Therefore they carry the thermodynamic information of the created matter directly.
This paper focuses on di-electron production in the low mass region below 1~GeV/$c^{2}$ 
and covers the following two topics.
The first topic is the measurement of the di-electron yield in the low-$p_{T}$ region,
which is considered to be an optimal window for detecting thermal $q\bar{q}$ and $\pi\pi$ 
annihilations.
The second one is the measurement of the direct photon yield with the virtual photon 
method ($\gamma^{\ast} \rightarrow e^{+}e^{-}$).
Thermal photons from QGP are considered to be a primary contributor in 
$1.0 < p_{T} <  5.0~$GeV/$c$ at RHIC energy~\cite{PRC1}.

\section{Low-mass Low-$p_{T}$ Di-electron Continuum}
\begin{figure}[htb]
\center{\psfig{figure=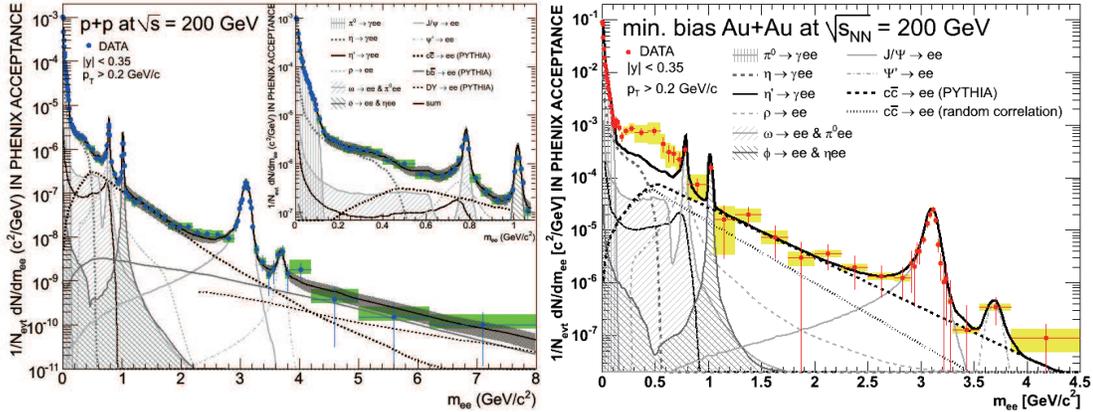,height=2.2in}}
\caption{Invariant mass spectrum for p+p (left) and Au+Au (right) collisions compared 
to the hadronic cocktail calculations.
\label{fig:Invm}}
\end{figure}

The e$^{+}$e$^{-}$ pair yield was measured in p+p and Au+Au collisions at PHENIX and 
hadron decay components were estimated using a hadronic cocktail calculation which 
incorporated the measured yields of the mesons for both p+p and Au+Au 
collisions~\cite{PH1}$^{,}$~\cite{PH2}.
The left and right panels in Fig.~\ref{fig:Invm} show the invariant mass spectra 
in p+p and Au+Au collisions compared to the hadronic cocktail calculations.
The symbols and lines indicate the real data and the cocktail calculations, 
respectively. 
While the real data is in excellent agreement with the cocktail calculation in p+p 
collisions, a large enhancement over the calculation is clearly seen in 
$150 < m_{ee} < 750~$MeV/$c^{2}$ in Au+Au collisions.
The ratio of the integrated yield in $150 < m_{ee} < 750~$MeV/$c^{2}$ between the real and 
the cocktail calculation is 3.4$\pm$0.2(stat)$\pm$1.3(sys)$\pm$0.7(model).
This enhancement is prominent in $p_{T} < 1.0~$GeV/$c$~\cite{AT}.

\section{Low $p_{T}$ Direct Photon}
In general, any source of real photons can emit virtual photons which convert to low 
mass e$^{+}$e$^{-}$ pairs.
A direct photon production process such as $q + g \rightarrow g + \gamma$ has an 
associated process that produces low mass e$^{+}$e$^{-}$ pairs, i.e. 
$q + g \rightarrow g + \gamma^{\ast} \rightarrow q + e^{+}e^{-}$.
The relation between the photon production and the associated e$^{+}$e$^{-}$ pair 
production is expressed by the Kroll-Wada formula, Eq.~\ref{eq:Kroll}~\cite{KW}.
\begin{equation}
\frac{d^{2}n_{ee}}{dm_{ee}} = \frac{2\alpha}{3\pi} \frac{1}{m_{ee}} 
\sqrt{1-\frac{4m_{e}^{2}}{m_{ee}^{2}}} \left( 1+\frac{2m_{e}^{2}}{m_{ee}^{2}} \right) 
S dn_{\gamma},
\label{eq:Kroll}
\end{equation}
where $\alpha$ is the fine structure constant, $m_{e}$ and $m_{ee}$ are the masses of the 
electron and the e$^{+}$e$^{-}$ pair, respectively, and $S$ is a process dependent 
factor.

In the case of hadrons such as $\pi^{0}$ and $\eta$ Dalitz decays, $S$ is given as 
$S = |F(m_{ee}^{2})|^{2} \left( 1- \frac{m_{ee}^{2}}{M_{hadron}^{2}} \right)^{3}$, 
where $F$ denotes the form factor and $M_{hadron}$ is the mass of the parent hadron.
The $S$ factor is zero for $m_{ee} > M_{hadron}$.
On the other hand, the $S$ factor becomes unity for $m_{ee} \ll p_{T}$ in the case of 
virtual direct photon.
Therefore it will be possible to extract the virtual direct photon component from the 
e$^{+}$e$^{-}$ pair mass spectra by utilizing the difference in e$^{+}$e$^{-}$ pair mass 
dependence of the $S$ factor.

\begin{figure}[htb]
\center{\psfig{figure=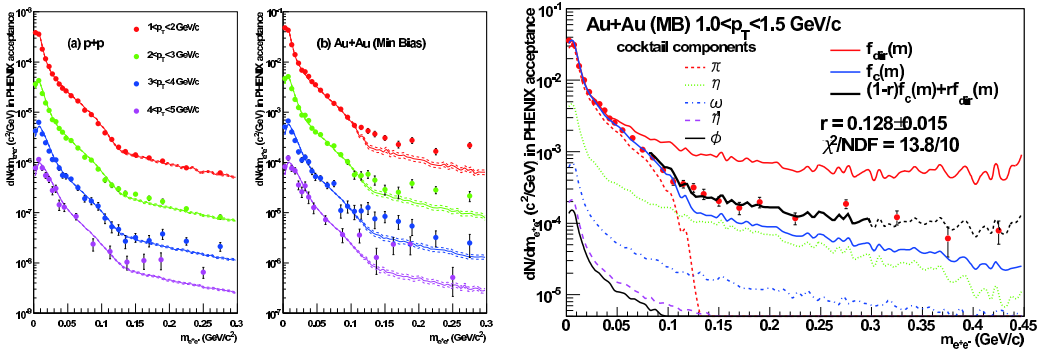,height=2.0in}}
\caption{The left and center panels show the invariant mass spectra in p+p and Au+Au 
  collisions for several $p_{T}$ bins compared to the cocktail calculations.
  The right panel shows the mass spectrum in Au+Au collisions for 
  $1.0 < p_{T} < 1.5~$GeV/$c$ together with a fit result by Eq.~\ref{eq:ffit}.
\label{fig:spec_fit}}
\end{figure}
An enhancement of the e$^{+}$e$^{-}$ pair yield for Au+Au collisions is clearly visible
in the $p_{T}$ region above 1~GeV/$c$ and the mass region of 
$100 < m_{ee} < 300~$MeV/$c^{2}$ as shown in the center panel of Fig.~\ref{fig:spec_fit}.
We would like to emphasize that a small excess over the cocktail is also observed for 
p+p collisions in the high $p_{T}$ region as shown in the left panel of 
Fig.~\ref{fig:spec_fit}.

The following function is used for fitting the data to determine the fraction of the virtual 
direct photon component in the mass spectrum.
\begin{equation}
f(m_{ee}) = (1-r) \cdot f_{cocktail}(m_{ee}) + r \cdot f_{direct}(m_{ee}),
\label{eq:ffit}
\end{equation}
where $f_{cocktail}$ is the mass distribution from the decay of neutral hadrons estimated 
using the cocktail calculation and $f_{direct}$ is that from the virtual direct photon decays, 
and $r$ is the virtual direct photon fraction.
Assuming that this excess comes from virtual direct photons, the fit to the result for 
Au+Au collisions in $1.0 < p_{T} < 1.5~$GeV/$c$ gives $\chi^{2}/NDF = 13.8/10$.
On the other hand, assuming that this excess comes from $\eta$, i.e. $f_{\eta}(m_{ee})$ 
is used in place of $f_{direct}(m_{ee})$, the fit gives $\chi^{2}/NDF = 21.1/10$.
The result favors the virtual direct photons.
It is also to be noted that ascribing this enhancement to $\eta$ requires anomalous $\eta$ 
enhancement, which seems very unlikely.

\begin{figure}[htb]
\center{\psfig{figure=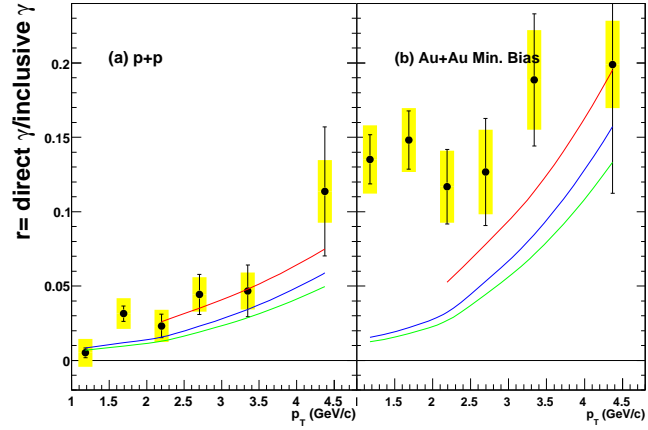,height=2.50in}}
\caption{The obtained fractions of the virtual direct photon component as a function of 
$p_{T}$ in p+p (left) and Au+Au (right) collisions.
\label{fig:r_gamma}}
\end{figure}
Figure~\ref{fig:r_gamma} shows the obtained fractions of the virtual direct photon 
component as a function of $p_{T}$ in p+p and Au+Au collisions.
The symbols show the result and the lines are the expectations from  
next-to-leading-order perturbative QCD (NLO pQCD) calculations~\cite{pQCD} with 
different theoretical scales.
A clear excess above the NLO pQCD calculation is seen in Au+Au collisions while the 
result in p+p collisions is consistent with the NLO pQCD calculation.

\begin{figure}[htb]
\center{\psfig{figure=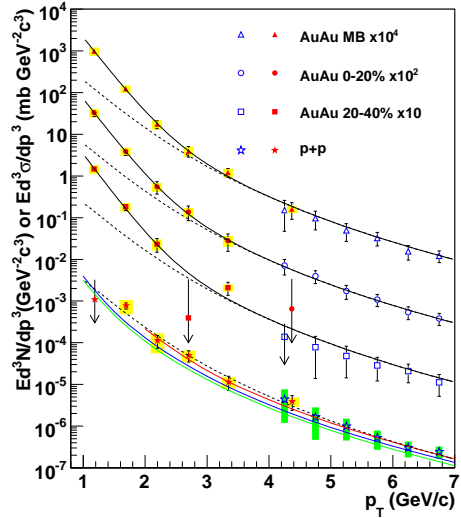,height=3.00in}}
\caption{The direct photon spectra in p+p and Au+Au collisions as a function of $p_{T}$.
\label{fig:spec_photon}}
\end{figure}

Finally, the real direct photon yield is obtained by multiplying the inclusive photon 
yield to the virtual direct photon fraction.
Figure~\ref{fig:spec_photon} shows the direct photon spectra in p+p and Au+Au collisions.
This is the first time that the direct photon production in p+p collisions has been 
measured in $1.0 < p_{T} < 4.0~$GeV/$c$.
The result in p+p collisions is well described by a modified power law function as 
shown by the dashed curve.
The obtained yield in Au+Au collisions is above the $N_{coll}$-scaled p+p fitted curve 
in the low $p_{T}$ region, where $N_{coll}$ is the number of collisions 
Fitting the result for Au+Au central collision with an exponential plus the $N_{coll}$-scaled 
p+p fitted curve provides the inverse slope parameter of $221 \pm 23 \pm 18~$MeV~\cite{PH3}.
Comparing to available theoretical models gives a handle on the initial temperature that 
ranges from 300 to 600~MeV with $\tau_{0} \sim $0.6-0.15~fm/$c$. 
Implications of the results are under various investigations.


\section*{References}
\begin{thebibliography}{99}
\bibitem{PRL1}K.~Adcox {\it et al}, \Journal{\PRL}{88}{022301}{2002}.

\bibitem{PRL2}K.~Adcox {\it et al}, \Journal{\PRL}{98}{172301}{2007}.

\bibitem{PRL3}S.S.~Adler {\it et al}, \Journal{\PRL}{91}{182301}{2003}.

\bibitem{PRC1}S.~Turbide {\it et al}, \Journal{{\em Phys. Rev.} C}{69}{014903}{2004}.

\bibitem{PH1} A.~Adare {\it et al}, arXiv:0802.0050 [nucl-ex].

\bibitem{PH2} S.~Afanasiev {\it et al}, arXiv:0706.3034 [nucl-ex].

\bibitem{AT} A.~Toia, arXiv:0805.0153 [nucl-ex].

\bibitem{KW}N.W.~Kroll and W.~Wada, \Journal{{\em Phys. Rev.}}{98}{1355}{1955}.

\bibitem{pQCD}L.E.~Gordon and W.~Vogelsang, \Journal{\PRD}{48}{3136}{1993} and 
  W.~Vogelsang, private communication.

\bibitem{PH3} A.~Adare {\it et al}, arXiv:0804.4168 [nucl-ex].






\end{thebibliography}
\end{document}